\documentstyle[12pt,aasms4]{article}

\slugcomment{\shortstack[r]{submitted to {\sl Astrophysical Journal Letters}
}}

\begin{document}

\title
{The Infrared Luminosity Function of Galaxies in the Coma Cluster}

\author
{Roberto De Propris\altaffilmark{1} and Peter R. Eisenhardt\altaffilmark{2}}

\affil
{MS 169-327, Jet Propulsion Laboratory, California Institute of Technology, 
Pasadena, CA, 91109; propris@coma.jpl.nasa.gov, prme@kromos.jpl.nasa.gov}

\altaffiltext{1}
{National Research Council Resident Research Associate}

\altaffiltext{2}
{Visiting Astronomer, Kitt Peak National Observatory, National Optical
Astronomy Obsevatories, which is operated by the Association of Universities
for Research in Astronomy, Inc., under cooperative agreement with the
National Science Foundation}

\author{S. Adam Stanford\altaffilmark{2}}

\affil
{Institute of Geophysics and Planetary Physics, Lawrence Livermore
National Laboratories, Livermore, CA, 94550; e-mail: adam@igpp.llnl.gov}

\and

\author{Mark Dickinson\altaffilmark{2}$^,$ \altaffilmark{3}}

\affil{Department of Physics and Astronomy, The Johns Hopkins University,
Baltimore, MD 21218; e-mail: med@stsci.edu}

\altaffiltext{3}{Allan C. Davis Fellow, also with the Space Telescope
Science Institute}
\authoremail{med@stsci.edu}


\begin{abstract}

An infrared survey of the central 650 arcmin$^2$ of the Coma cluster
is used to determine the $H$ band luminosity function for the cluster.
Redshifts are available for all galaxies in the survey with $H < 14.5$,
and for this sample we obtain a good fit to a Schechter function with
$H^*=11.13$ and $\alpha=-0.78$.  These luminosity
function parameters are similar to those measured for field galaxies
in the infrared, which is surprising considering the very different
environmental densities and, presumably, merger histories for field
galaxies. For fainter galaxies, we use two independent techniques
to correct for field galaxy contamination in the cluster population:
the $B-R$ color-magnitude relation and an estimate for the level of
background and foreground contamination from the literature.  Using 
either method we find a steep upturn for galaxies with $14 < H < 16$, 
with slope $\alpha \simeq 1.7$.

\end{abstract}


\section
{Introduction}

The availability of wide-field CCD detectors on large telescopes has
renewed interest in the luminosity functions of galaxies in clusters,
which can now be determined to $M_B \sim -10$ at 100 Mpc, a 
luminosity limit comparable to that formerly reached in the Local Group (e.g.,
van den Bergh 1992). Coma, as one of the nearest ($z = 0.023$)
and richest (R=2) Abell clusters is, together with Virgo, one of the
best studied systems (see Mazure et al.\ 1998 for a compendium of
recent work on the Coma cluster).

Optical luminosity functions for bright galaxies in Coma are consistent
with a Schechter (1976) luminosity function
having $\alpha \sim -1$ (e.g., Lopez-Cruz et al.\ 1997).  Dwarf galaxies
are better fit by a power-law with $\alpha = -1.4$ (in the Schechter
formalism).  These properties are similar to those of the luminosity
function in Virgo (Sandage, Binggeli \& Tammann 1985) and in Fornax
(Ferguson \& Sandage 1988).  Recently, however, Lobo et al.\ (1997a)
have claimed a steep $V$ band slope for faint galaxies in Coma ($\alpha
\sim -1.8$), supporting earlier claims for a large population of dwarf
galaxies in clusters (e.g., De Propris et al.\ 1995).  Table 1 summarizes
recent determinations of the faint-end slope of the luminosity function
in the Coma cluster.

It is generally difficult to relate optical luminosity functions to the
underlying mass distribution, owing to our incomplete understanding
of star formation in galaxies. It has been known for some time that
cluster galaxies have steeper LF's than in the field (Binggeli, Sandage
\& Tarenghi 1990), in contrast to Cold Dark Matter (CDM) models
which predict $\alpha=-2$ (e.g., White \& Frenk 1991).

On the other hand, a steep optical LF may be compatible with a flat 
mass function if dwarf galaxies have their luminosities boosted by
fading starbursts (Hogg \& Phinney 1997). There is indeed some evidence
for recent (1 -- 2 Gyr ago) star formation among dwarfs in Coma (Donas,
Milliard \& Laget 1995) and in Virgo and Fornax as well (Held \& Mould
1994 and references therein).

Infrared luminosities are known to be less sensitive to star formation
and to correlate well with dynamical mass (Gavazzi, Pierini \& Boselli
1996). Therefore infrared luminosity functions should be expected
to approximate better the underlying mass function.  A study of the
infrared luminosity function of Coma is timely because of the recent
publication of the first infrared-selected luminosity functions for
field galaxies (Gardner et al.\ 1997 - G97, Szokoly et al.\ 1998), allowing a
comparison of mass distributions in two environments of highly different
densities. While Mobasher \& Trentham (1998) have recently presented a
$K$-band study of faint galaxies in Coma, their survey covers only 41
arcmin$^2$ and hence the luminosity function is not well constrained.

Eisenhardt et al.\ (1998 - hereafter EDGSWD) have obtained photometry in
$U,B,V,R,I,z,J,H$ and $K_s$ for $\sim 500$ infrared selected galaxies
in a $29.2' \times 22.5'$ field in the core of the Coma cluster.
EDGSWD selected objects at $H$ (and required confirmation at $J$)
to provide a baseline for comparison to $K$ selected samples in more
distant clusters (Stanford, Eisenhardt \& Dickinson 1995, 1998), and
also because the $H$ data reached approximately 0.5 mag deeper.  Here we
use these data to derive the $H$ luminosity function for galaxies in the
Coma cluster.  The conversion from $H$ to $K_s$ is provided in \S 3.1.
A detailed description of the data, observations, data reduction and
photometry is given in EDGSWD, and is not repeated here, except for some
essential points.  We assume a redshift of 6950 km s$^{-1}$ (Mazure \&
Gurzadyan 1998).

\section
{The Luminosity Function}

Star-galaxy separation was determined in the $R$ band using the Kron
(1980) $r_{-2}$ parameter (EDGSWD). Because the available membership
information is qualitatively different for bright vs. faint galaxies,
we consider them separately. For galaxies with $H < 14.5$ redshifts
are available in the literature (Lobo et al.\ 1997b). All galaxies
with $3000 < cz < 10000$ km/s are assumed to be members and at the same
distance. For fainter galaxies ($14.0 < H < 16$) we use two independent
methods to correct for non-cluster members. Following Mazure et al.\
(1988) and Biviano et al.\ (1995), we assume that galaxies within $\pm
0.3$ magnitudes of the $B-R$ color-magnitude relation defined by the
brighter early type galaxies are members. The relation and the color
selection criterion are shown in Figure 1.  We also use the infrared
counts by Huang et al.\ (1997), tranformed to the $H$ band using our
$H-K$ color magnitude relation, to estimate the amount of contamination
from background and foreground galaxies.

Galaxies are then counted in 0.5 magnitude bins. The resulting luminosity
function is shown in Figure 2.

\subsection{Bright Galaxies}
For bright galaxies we choose to fit a Schechter function, using the
maximum likelihood technique of Sandage, Tammann \& Yahil (1979).  The best
fitting parameters are $H^*=11.13$ and $\alpha=-0.78$. The inset in Figure
2 shows the $1\sigma$ error ellipse. The three brightest objects are
excluded from the fit, since we are unable to fairly sample their space
density and such objects fall outside of the extrapolation of the Schechter
function at bright magnitudes (Schechter 1976).

To illustrate the uncertainty in $H^*$ and $\alpha$, 1000 Monte Carlo
simulations were generated in which 111 objects were drawn from a
Schechter function whose parameters were the best fit to our `original'
data and with errors as described in EDGSWD. These artificial data were
fitted using the same method as above, and the results are shown in the
inset in Figure 2.

\subsection{Faint Galaxies}

As shown in Figure 2, both the color selection method and the
background subtraction method result in a significant excess of faint
galaxies relative to the Schechter function for bright galaxies
determined above. Fitting to the $B-R$ selected counts for $14 < H < 16$
we find $\alpha=-1.73 \pm 0.14$. Could this slope be an artifact ?

Magnitude errors and spurious detections can cause an artificial
steepening in the derived slope (Kron 1980).  However, the estimated
errors are smaller than the width of the bin even for the faintest
galaxies considered, and spurious detections should be few because
objects were required to be detected in both the $J$ and $H$ images.  
In fact we are quite likely to have {\it underestimated} the
number of dwarfs due to the difficulty in detecting low surface
brightness galaxies against the bright infrared background.  The
completeness is estimated from Monte-Carlo simulations to be $>80\%$ to
$H=16$ for galaxies with $r_1 < 1.4 h^{-1}$ kpc and $> 90\%$ to
$H=16.5$ for $r_1 < 0.7 h^{-1}$ kpc (EDGSWD), where $r_1$ is the first
moment of the light distribution (Kron 1980).  Because Fornax cluster
dwarfs typically have effective radii $< 1 h^{-1}$ kpc (Ferguson 1989),
we do not expect this to be a large correction.

Misclassifying stars as galaxies cannot account for the upturn: there
are no significant discrepancies in star/galaxy classification between
EDGSWD and Lobo et al. (1997a)

A more serious concern is that the slope of the infrared background
counts is similarly steep (equivalent to $\alpha = -2.7$, Huang et al.
1997), raising the possibility that an incorrect removal of contaminating
objects is responsible for the upturn.  We consider this unlikely because
(i) counts obtained via color selection agree well with those derived via
background subtraction techniques; a narrower color-selection strip ($\pm
0.15$) yields a slightly flatter slope ($\alpha=-1.55 \pm 0.20$) but the
upturn remains significant; (ii) the excess required to account for our
result is a factor of 2.5 above the estimated background counts; and (iii)
galaxy counts in the direction of the Coma cluster are found to be in
satisfactory agreement with those in the general field (Secker \& Harris
1996).

Nevertheless, the question of cluster membership for these galaxies is the
dominant uncertainty in our measurement of the faint end slope, and we are
planning on a redshift survey of the faint sample, in order to address the
issue of their membership. Time has already been awarded on the WIYN
telescope to pursue this investigation.

\section{Discussion}

\subsection{Bright Galaxies}

The infrared luminosity function derived for bright galaxies ($H < H^*+3$)
is in good agreement with that at $R$ (Lopez-Cruz et al.\ 1997) as well
as at $V$ (Lobo et al.\ 1997a): all are reasonably well matched by a
Schechter function with a `flat' ($\alpha\sim -1$) power law.  This argues
for a relatively small contribution from young stars in these galaxies.

Our value of $H^*$ is in excellent agreement with the $M^*_K=-23.12 + 5 \log
(h)$ value reported by G97, using $H-K_s=0.22$ from EDGSWD and a 
$k$-correction of 0.05. Both this method and a direct determination 
from the $K_s$ data of EDGSWD, to the $M_K=-21 + 5 \log (h)$ limit
of G97, yield $M^*_K=-23.25 + 5 \log (h)$. The observed $B-H=4.2$
yields $M^*_B \approx -19.1 + 5 \log(h)$, vs. $M^*_B = -19.0 + 5 \log(h)$ 
for the Virgo cluster (Sandage et al.\ 1985). The agreement is somewhat 
fortuitous given the uncertainty in $H^*$ due to our smaller sample and
the fact that we do not properly survey the brightest galaxies. Our
bright galaxy slope ($\alpha=-0.78$) agrees well with G97's $\alpha=-0.91$
but depends on the cutoff magnitude. Using the G97 cutoff gives $\alpha
=-0.93$ in $H$ and $\alpha=-0.98$ from the $K_s$ data, suggesting that
our results may be influenced by the presence of the `dip' in the optical
luminosity function of Coma at $V\sim 17$ (Lobo et al.\ 1997a), equivalent
to $H \sim 14$.

The IR luminosity functions (and hence mass functions) of bright
galaxies in the field and in this rich cluster appear to be similar,
despite the roughly thousand-fold difference in environmental density.
This agreement is surprising because of the different morphological mixes
and, presumably, merger histories for these galaxies.  If mass is truly
the defining parameter in controlling the bulk properties of galaxies
and their morphology (see Gavazzi et al.\ 1996) this similarity supports
models in which large galaxies form at high redshift and evolve passively
to the present epoch, and in which mergers are relatively unimportant
(a scenario also favored by Stanford et al. 1998).

\subsection{Faint Galaxies}

For galaxies fainter than $H=14.5$, we find a steep upturn in the
luminosity function ($\alpha\approx -1.7$). Due to the uncertainty in
cluster membership, this result should be considered to support claims
for a steep luminosity function for dwarfs, rather than providing a
precise estimate of the slope of the faint end of the luminosity
function.  Nevertheless, we detect a population of dwarfs
about two times larger than expected from the $\alpha=-1.4$ found
by many previous authors (Table 1).

Because $H$ luminosity is linearly correlated with mass for field and
Virgo cluster galaxies (Gavazzi et al. 1996), the most natural
interpretation of the steep infrared slope is that it represents a real
increase in the space density of low mass galaxies in Coma, rather than
an enhanced star formation rate in such objects.

There is some suggestion in Table 1 that the luminosity function slope
increases with clustercentric radius (e.g. Lobo et al. 1997a).  Such a
trend might be caused by a higher incidence of mergers or tidal
disruption in the dense cluster core.   On the other hand dwarfs may
actually {\it form} in mergers (Krivitsky \& Kontorovich 1997) and in
tidal tails (Hunsberger, Charlton \& Zaritsky 1996).  The reality of
the trend towards steeper slopes at larger radii remains inconclusive:
Lopez-Cruz et al. (1997) and Secker et al.\ (1997) find $\alpha=-1.4$
at $R$ in fields of similar size to ours, identical to the value found
at $R$ by Bernstein et al. (1995) in a small field near the cluster
center.

Another possibility is that most of the luminosity from dwarf galaxies
comes from fading bursts of star formation, leading to steeper faint
end slopes at longer wavelengths, as predicted by Hogg \& Phinney
(1997).  Because the burst luminosity becomes fainter and redder with
time, there is an increased probability of finding faint, red
galaxies.  Given the difference in mass to (near infrared) light ratio
of a $10^8$ vs. a $10^{10}$ year old population (Bruzual \& Charlot
1993), starbursts producing $\sim10\%$ of the mass in the underlying
population every few hundred million years would satisfy the
requirements of the Hogg \& Phinney model.  Although some Fornax
cluster dwarfs show evidence for a substantial young population (Held
\& Mould 1994), none of the Local Group dwarfs, with the possible
exception of the Fornax dwarf (Gallagher \& Wyse 1994) show such
evidence.  Again, existing data on the correlation of slope with
wavelength are inconclusive (Table 1).

It is also plausible that the dwarf galaxy H luminosity function is more
accurately a Schechter function than a power law, and the steep slope
we measure is only an approximation to the exponential portion of the
function.  Binggeli, Sandage \& Tammann (1988) suggest $M^*_B = -15.9 + 5
\log(h)$ for dwarfs, corresponding to $H \approx 15.5$ in Coma.  In this
case the data of Mobasher \& Trentham (1998), which samples a smaller
field to $K = 18.5$, provides a better estimate of the faint end slope.

A comparison with field dwarfs is difficult, since no infrared survey
has yet reached luminosity limits as deep as ours (which is equivalent
to $M_K \sim -18.5 + 5 \log(h))$. Using a $K_s$-selected sample with 
110 redshifts, Szokoly et al.\ (1998) do find a
steeper slope ($\alpha \sim -1.3$) in their field infrared luminosity
function than G97, despite the similar luminosity
limit ($M_K \sim -21 + 5 \log(h)$) of the samples analyzed, but we
consider G97's result more reliable as it is based on
$\sim 500$ redshifts.  If the similarity between the IR and optical luminosity
functions in Coma also holds true in the field, the very steep slope
($\alpha \sim -2.8$) found by Loveday (1997) at $b$ for faint galaxies
in the Stromlo-APM survey may foretell an upturn in the IR field
luminosity function as well.  The 2MASS survey should settle this
issue, as it will reach $M_K \sim -16.5 + 5 \log(h)$ at the distance of
the Virgo cluster (Skrutskie et al. 1997).

Our results for Coma support the existence of an universal galaxy
luminosity function, which is well approximated by a flat Schechter
function for bright galaxies and a steep power-law for dwarfs (Trentham
1998b).  As demonstrated by EDGSWD and Skrutskie et al. (1997), it is now
possible to obtain `panoramic' data in the infrared: Other clusters
should now be studied with the same methods, in order to determine the
mass function of galaxies in clusters and study the effects of their
environment.

\acknowledgments

We would like to thank C. Pritchet for his help with Kron photometry
and some helpful comments, C. Lobo for providing her catalog ahead
of publication and J. Secker for some enlightening discussions. We
are grateful to the referee, J. P. Gardner, for a number of helpful
suggestions.  R. D. P.'s work is supported by the National Research
Council under its Resident Research Associateship program.  The research
described here was carried out by the Jet Propulsion Laboratory,
California Institute of Technology, under a contract with NASA.

\clearpage
\centerline{FIGURE CAPTIONS}
Figure 1 --- $B-R$ vs. $H$ color magnitude diagram for galaxies in the
field of the Coma cluster.

Figure 2 --- The luminosity function of the Coma cluster and the best
fitting Schechter function (solid line). 
Error bars for bright galaxies and for color-selected counts are assumed
to be Poissonian. For background selected counts we assume $\sqrt N$ errors
in the `raw counts' and add these in quadrature with errors in the 
estimated level of foreground and background contamination, evaluated 
according to equation (5) of Huang et al.\ (1997). The
differences between the values of $H^*$ and $\alpha$ retrieved from Monte 
Carlo simulations and the best fit values are plotted in the inset,
together with the $1\sigma$ error ellipse.
\clearpage

\begin{deluxetable}{llll}
\tablewidth{33pc}
\tablecaption{Recent Measurements of the slope of the Faint End of the Coma LF}
\tablehead{
\colhead{Field Size ($\sq '$)}           & \colhead{$\alpha$}      &
\colhead{Mag. Range}          & \colhead{Reference}  
}
\startdata
14300 & $-1.32$ & $ 17.0 < b < 20.0 $ &1 \nl
1200 & $-1.3 \pm 0.1 $ & $ 13.0 < b < 20.0 $ & 2 \nl
51 & $-1.51 \pm 0.13$ & $16.5 < V < 21.0$ & 3 \nl
1500 & $-1.80 \pm 0.05$ & $13.5 < V < 21.0$ & 3 \nl
674 & $\sim -1.7$ & $ 20.0 < R < 24.0 $ & 4 \nl
52 & $-1.42 \pm 0.05$ & $15.5 < R < 23.5$ & 5 \nl
529 & $-1.42 \pm 0.12$ & $ R < 22.5 $ & 6 \nl
700 & $-1.41 \pm 0.05$ & $15.5 < R < 22.5 $ & 7 \nl
657 & $-1.73 \pm 0.14$ & $ 14.0 < H < 16.0 $ & 8 \nl
41 & $-1.41 \pm 0.35$ & $15.5 < K < 18.5$ & 9 \nl

\tablerefs{
(1) Thompson \& Gregory 1993;
(2) Biviano et al. 1995;
(3) Lobo et al.\ (1997a); 
(4) Trentham 1998;
(5) Bernstein et al.\ 1995;
(6) Lopez-Cruz et al. 1997;
(7) Secker, Harris, \& Plummer 1997;
(8) this work; 
(9) Mobasher \& Trentham 1998 }
\enddata
\end{deluxetable}
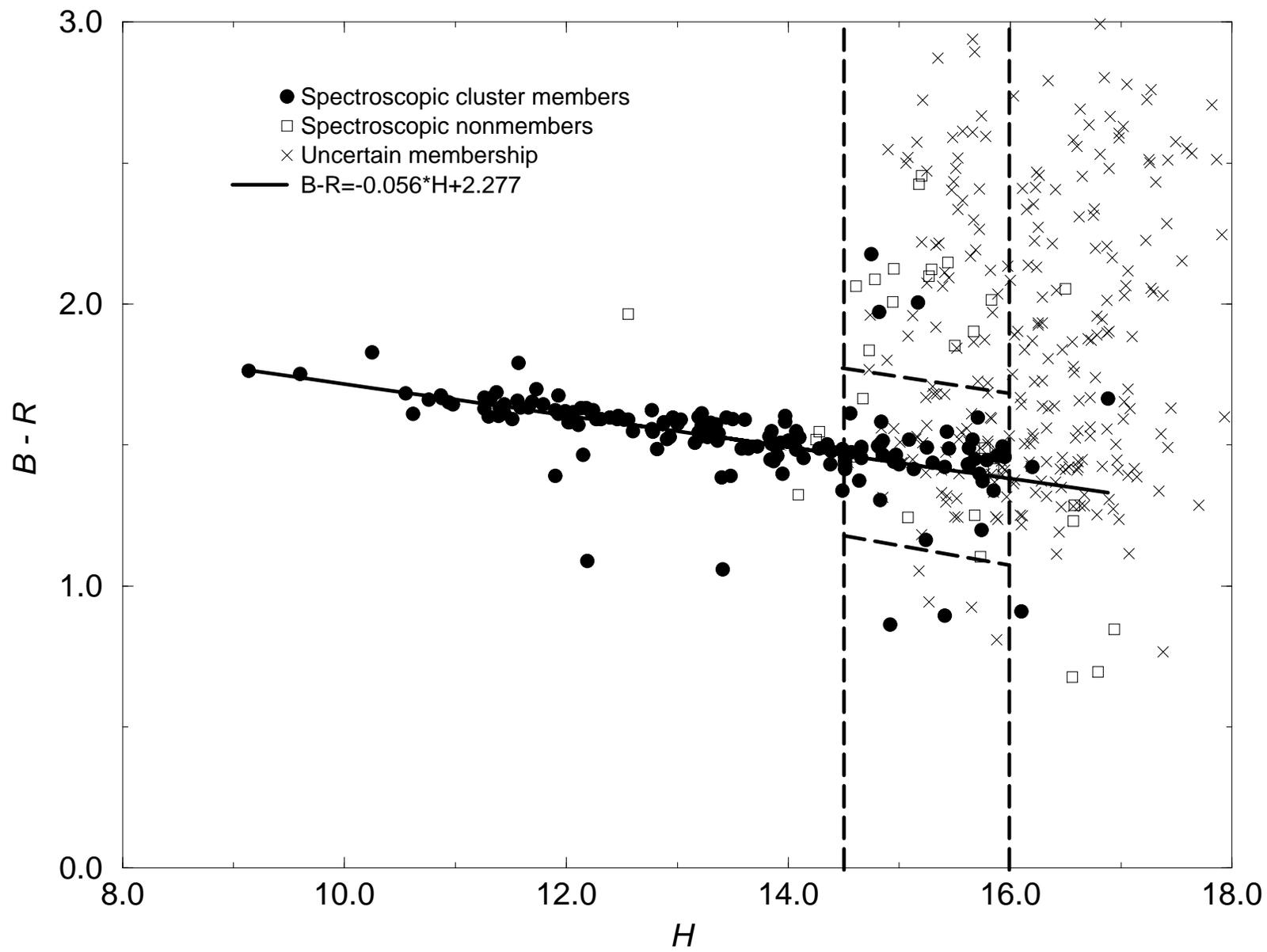
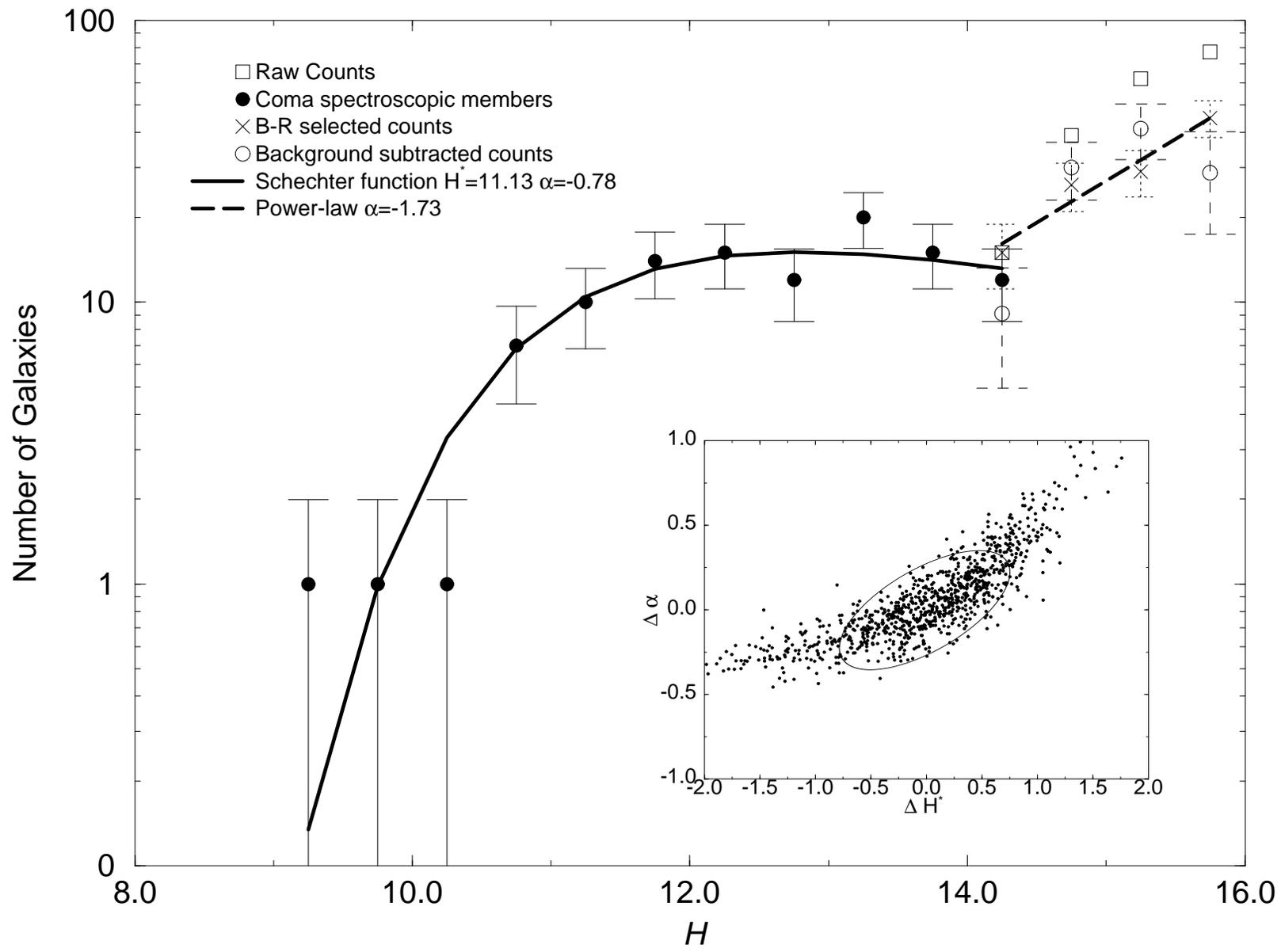
\end{document}